\documentstyle[sprocl]{article} 

\def\npb{{\em Nucl. Phys.} B} 
\def\plb{{\em Phys. Lett.}  B} 
 
\def\prd{{\em Phys. Rev.} D} 
 
\def\be{\begin{equation}} 
\def\ee{\end{equation}} 
\def\beq{\begin{equation}} 
 
\def\ee{\end{equation}} 
\def\bea{\begin{eqnarray}} 
\def\eea{\end{eqnarray}} 
 \def\ci{\cite}
 
\begin{document} 
 
\title{CONSTANT VERSUS FIELD DEPENDENT GAUGE COUPLINGS IN SUPERSYMMETRIC 
THEORIES\footnote{ 
Based on the 
talk given by F.Q.  at the {\it Phenomenological Aspects of Superstring 
Theories} (PAST97) Conference, ICTP Trieste, Italy, October 2-4 1997. Preprint 
IFUNAM FT98-4.}}

\author{C.P. Burgess$^a$, A. de la Macorra$^b$, I. Maksymyk$^c$ and F. Quevedo$^b$ } 

 \vskip 1cm

\address{$^a$ Physics Department, McGill University, 3600 University
 Street,Montreal, Quebec, H3A 2T8, 
Canada\vskip .1cm 
$^b$ Instituto de F\'{\i}sica, Universidad Nacional Aut\'onoma de M\'exico,
 Apartado Postal 20-364, 01000 
M\'exico D.F., M\'exico \vskip .1cm
$^c$ TRIUMF, 4004 Wesbrook Mall, Vancouver, BC, V6T 2A3, Canada.} 
 
 
\vskip 1cm

\maketitle 
\abstracts{We briefly discuss the differences between considering the gauge coupling as a constant or as a 
field, the dilaton,   
 in $N=1$ supersymmetric theories. We emphasize the differences regarding supersymmetry breaking. Recent 
developments on the nonperturbative dynamics of these theories provide new ideas on the induced dilaton 
potential and its stabilization.}

Thanks to the work of Seiberg and many others, the understanding of supersymmetric theories has improved 
considerably during the past three years \ci{review}$^{-}$\ci{seiberg}. 
In particular, the nonperturbative dynamics determining the possible 
 phases of these theories has been very well understood. This is important 
 for understanding issues such as chiral symmetry breaking, supersymmetry 
 breaking, the vacuum structure etc. In the context of string theory, 
 having $N=1$ supersymmetric theories as their low energy effective 
 theories, this 
 progress should be reflected on the possibility to address the most 
 important obstacles for the theory to make contact with low energy physics, 
namely, lifting the vacuum degeneracy and breaking supersymmetry. 
 
Superstring theories include always in their spectrum a massless field 
 called the dilaton $S$ which provides the bare gauge coupling. 
It is not only massless but has an exactly  flat potential in perturbation 
theory, and therefore it is one of the many `moduli' of the theory. 
 Nonperturbative effects generically lift this potential and the dilaton 
will get a mass. Depending on the nature of these effects, the mass of the 
 dilaton will be determined by the supersymmetry breaking scale and then is 
 expected to be small, or is fixed at the Planck scale and therefore the 
 dilaton does 
 not appear in the low-energy spectrum of the theory. In the first scenario 
the nonperturbative effect responsible for breaking supersymmetry is the 
 same that fixes the dilaton whereas the second scenario is in two steps: 
 a Planck scale effect fixes the dilaton and a low energy effect breaks 
 supersymmetry. 
These two different scenarios will generally be differentiated in the 
low energy action by having the gauge coupling either constant or field 
 dependent. The two-steps scenario fits with the gauge mediated 
 supersymmetry breaking scenario recently revived \ci{int}$^-$\ci{exampletwo}
whereas the one-step 
 scenario fits with the more standard gravity mediated supersymmetry 
 breaking scenario. 
 
The recent studies in supersymmetric gauge theories consider 
 the gauge coupling as a constant and therefore it applies to the 
 two-steps scenario directly. The search for models that break 
 supersymmetry therefore has a direct application to string theory, 
 only on this scenario. It is then interesting to ask what are  
the implications of the new understanding of supersymmetric gauge  
theories for the more standard scenario where the dilaton survives at 
 low energies. 
 
Previous attempts to understand these issues were based on a limited 
 knowledge of supersymmetric theories, and the typical models considered 
 included  
a string hidden sector with several gauge group factors and matter charged 
 under one of the factors \ci{review}. Concentrating on the dilaton field, the standard 
superpotentials that emerge in this case are of the form \ci{ccm}$^,$\ci{racetrack}
\begin{equation} 
W=\Sigma_i A_i \; e^{-a_i S} 
\label{W1}\end{equation} 
These models generally do not break supersymmetry in the $S$ sector, 
they have a supersymmetric minimum at finite values of $S$, but  also 
have a runaway solution to zero coupling ($S\rightarrow \infty$). 
This behaviour at infinity has been argued on very general grounds by Dine 
 and Seiberg some time ago \ci{genericprob}. They argue that at zero coupling the theory 
 should be free and therefore the potential should vanish there. This is a 
 source of a cosmological problem pointed out by Brustein and Steinhardt \ci{BS}. 
Taking the superpotential above, being so steep, if the dilaton field starts  
at any value, it may never end up at the local minimum with non-vanishing 
 coupling but will roll all the way to the runaway vacuum.  
 
On the other hand, besides the field $S$, there is usually another  
modulus,  the field $T$ measuring the size of the compact space. 
This field has also a flat potential to all orders in perturbation theory 
that gets lifted by nonperturbative effects. The properties of $T$ and $S$ 
 are very similar and this similarity was actually at the origin of the 
 proposal of $S$ duality, given that there existed a better established 
 $T$ duality. 
There are even some models that have the symmetry $S\leftrightarrow T$. 
In the same way that $S$ represents the string coupling, $T$ represents the 
coupling of the underlying $2D$ sigma model. Curiously the potential for 
 the $T$ field found in simple examples, blows up for large values of $T$. 
Unlike what it was naively expected, that it should runaway to the 
weak coupling limit $T\rightarrow\infty$. This may  be understood in the 
 following way: 
 the gauge coupling in $10D$, $g_{10}$ is related to 
 the gauge coupling in $4D$, $g_4$ by $1/g_4^2=R^6/g_{10}^2$ where $R$ is 
 the size of the compact $6D$ space and so it is the real part 
 of the field $T$. 
A large value of $T$ is a large value of $R$, combined with a relatively  
small value of $g_4$ implies a very strong string coupling in $10D$,  
therefore 
the blowing-up of the potential is a strong string coupling effect,  
not controlled in string perturbation theory where the calculation 
 was performed. 
 
On the other hand the potentials for the $S$ field seem to behave very 
 different from those for the $T$ field. 
 It is then valid to question the general  
assumption that the potential for the $S$ field runs away to $\infty$ 
and study different alternatives to the sum of negative exponentials of 
 the equation above. 
 
  In this talk we will present several models illustrating the difference 
 between constant and dilaton dependent gauge couplings as well as different   
examples where the dilaton potential does not runaway to infinity. 
We  also argue that the inclusion of field-dependent gauge 
couplings can qualitatively change whether or not a given model 
spontaneously breaks supersymmetry. 
The main difference is due to the additional requirement of extremizing 
the superpotential with respect to the coupling-constant field. 
For instance, it can happen that a supersymmetry-breaking ground 
state for fixed gauge coupling becomes supersymmetric 
once the coupling constant is allowed to relax to minimize the 
energy. In particular we show that most of the models with dynamical supersymmetry 
 breaking, when the gauge coupling is field dependent, 
 do not break supersymmetry. Furthermore,  we find that the opposite of this is also possible,
supersymmetry can be unbroken for fixed gauge coupling, 
but  breaks down  once the gauge coupling is considered as a field. 

An example on the difference between having field  dependent or independent gauge couplings
is  the simplest case of gaugino condensation for a pure gauge theory 
having a simple gauge group and no matter multiplets \ci{review}. In this case, 
for constant gauge couplings, gauginos condense without breaking 
supersymmetry \ci{vy}.  The reason is that 
the gaugino condensate is given as the lowest component of a chiral superfield $U=<\lambda \lambda>$ and a 
non-vanishing value for the lowest component does not break supersymmetry.   
On the other hand, once a field dependent coupling constant is introduced via a chiral field $S$, whose real 
part  gives the coupling constant $Re\,S=1/g^2$,  a non-vanishing  gaugino condensate will break 
supersymmetry because it will be  proportional to the $F$ term of the $S$ field, and a  nonvanishing $F$ term 
breaks supersymmetry.
However, the dynamics of the dilaton field for a single gaugino condensate in pure Yang-Mills theory has a 
runaway behaviour $S\rightarrow \infty$ and the gaugino condensate vanishes $U=const. e^{- a S} 
\rightarrow 0$. Therefore in both cases, field dependent and field independent  coupling 
constant, supersymmetry is not broken. But in the fist case gauginos condense
whereas in the second case they do not condense.

 We will now study  potentials including matter fields  and we will consider
 $N=1$ supersymmetric models with gauge group 
$SU(N_c) $. 
We represent the matter multiplets with chiral superfields, 
$Q^i_\alpha \in R$ (and $\tilde{Q}_i^\alpha \in \tilde{R}$ ), where `$i$' is the flavour 
index, and `$\alpha$' is the gauge index. The 
kinetic microscopic action for 
the model is given  by $
\it{L}_{\rm kin} =  \frac{1}{4} f \, Tr W_{\alpha}W_{\alpha}
$, where $f$ is the gauge kinetic function and $W_{\alpha}$ the chiral gauge superfield  and we take standard
  kinetic terms for the matter supermultiplets. At tree level in string theory on has $f=k\, S$ with $k$ the Kac-Moody level.  
The microscopic superpotential relating 
the matter supermultiplets 
is taken to vanish identically, $W(Q,\tilde{Q}) = 0$.   

 To determine  the superpotential 
for the quantum `effective 
action' which generates the irreducible correlation functions 
of the theory (as opposed, say, to the theory's Wilson action) we study the operators 
whose correlations we wish to explore.
 Of particular interest, however, are those fields which can 
describe the very light scalar degrees of freedom of the model, since 
these describe the system's vacuum moduli and symmetries. 
In the absence of a microscopic superpotential for the matter fields 
$Q$ and $\tilde{Q}$, these light degrees of freedom are described 
classically (and hence also to all orders of perturbation theory) 
by the $D$-flat directions, which parametrize the 
zeroes of the classical scalar potential. 
It is well known that these $D$-flat directions 
can be parametrized in terms of a suitably chosen set of 
gauge-invariant holomorphic polynomials \ci{gaugeinvariants,pr}. 
We take  the arguments of the superpotential to be $W(U, M^i_j)$, where  $M^i_j = <Q^i_\alpha 
\tilde{Q}_j^\alpha >$,  $U  = < Tr W_{\alpha}W_{\alpha} >$. 
  Although the gaugino condensate 
field, $U$, does {\it not} similarly describe a $D$-flat direction, it is 
nonetheless convenient to keep it as an argument of the 
effective action.  
 The  superpotential  is completely determined by the twin conditions 
of linearity and symmetry under the model's global 
flavour symmetries.  As was demonstrated in \ci{bdqq}, the fact that 
$S$ only couples to the microscopic theory 
{\it via} the kinetic term implies, as an exact result, 
that the effective 
superpotential necessarily has the form \ci{tvy,kl}
 \be 
W = \frac{1}{4}\, US  + f(U, M^i_j) \; . 
\ee  
That is, $S$ can only appear linearly, and moreover only in the 
term $\frac{1}{4} US$.  Second,
the function $f(U, M^i_j)$ is determined by the various global 
chiral symmetries of the underlying supersymmetric gauge 
theory. In the absence of a superpotential for the matter 
fields, $Q^i_\alpha$ and $\tilde Q^\alpha_i$, the underlying 
gauge theory admits the classical global symmetry 
$SU(N_f)_L \times SU(N_f)_R \times U(1)_A \times U(1)_B \times U(1)_R$, 
of which the factors $U(1)_A \times U(1)_R$ are anomalous. 
Invariance of the effective superpotential 
under the anomaly-free symmetries implies the fields $M^i_j$ 
can appear only through the invariant combination $\det M$. 
(For $N_c < N_f$ we imagine the expectation value of the  
baryon operator, $B^{i_1\cdots i_{N_c}} = \epsilon^{\alpha_1 
\cdots \alpha_{N_c}} \, {Q^{i_1}}_{\alpha_1} \cdots
  {Q^{i_{N_c}}}_{\alpha_{N_c}}$ 
to be minimized by zero).  
The two anomalous symmetries, $U(1)_A$ and $U(1)_R$, 
then fix the form of the unknown function $f(U, \det M)$.

{}From these considerations, 
it is clear that $W$  has  the general structure   
 \be 
W = \frac{1}{4} US + \frac{U}{32 \pi^2 } 
\left[ (N_c-N_f) \log\left( { U \over \mu^3} \right) + 
 \log\left( { \det M \over \mu^{2 N_f} }\right) 
+ C_0 \right].
\label{W2}\ee  
 Symmetry arguments 
cannot determine the constants $\mu$ and $C_0$. 
Indeed $C_0$ may be chosen to vanish through an appropriate 
choice for $\mu$.

Since $W$ is the superpotential for the effective action --- as 
opposed to the Wilson action --- the correct procedure 
for `integrating out' fields is to remove them by solving their 
extremal equations for $W$, rather than by performing their 
path integral. Furthermore, for supersymmetric theories 
this should be done using the effective superpotential, $W$, rather 
than the effective scalar potential $V$.
Performing this operation for the gaugino condensate $U$ one obtains 
\be
W=c \,\left( \frac{\mu^{3N_c-N_f}e^{-8\pi^2 S }}{ \det M  }\right)^{1/(N_c-N_f)}=  c' \,\left( 
\frac{\Lambda^{3N_c-N_f} }{ \det M  } 
\right)^{1/(N_c-N_f)}
\label{W}\ee
where $c= - \; \frac{a}{32\pi^2} \,\exp\left(\frac{C_0+a}{a}\right)$, 
$c'=c\, \exp\left(-\frac{C_0}{a}\right)$ and the second equality 
defines the RG-invariant scale, $\Lambda=\mu^{3N_c-N_f} e^{-8\pi^2 S/(3N_c-N_f)}$.

 It is convenient to 
distinguish four different cases depending on the matter content: i)  $N_f < N_c$ ,
ii)  $N_f=N_c $, iii) $N_f > N_c$ and iv) $N_f > 3 N_c$.

In the first case, $N_f<N_c$,   the only invariant are meson fields and a non-vanishing superpotential $W= 
c (\frac{\Lambda ^{3N_c-N_f}}{det M})^{1/(N_c-N_f)}$ is obtained. Since the scale $\Lambda$ 
in terms of the coupling constant is given  by $\Lambda = \mu^{3N_c-N_f} e^{- 8\pi^2 S/(3N_c-N_f)}$  
minimizing the superpotential with respect to $S$ gives a runaway behaviour $S \rightarrow \infty$ and $W_S 
\propto W 
\propto e^{- 8\pi^2 S/(N_c-N_f)} \rightarrow 0$.  As in the pure Yang-Mills case, a   superpotential 
is dynamically generated
  but  its minimum is at vanishing potential and a supersymmetric vacuum is obtained.  However, for  a    field 
independent  gauge coupling we do not extremize the superpotential
with respect to $S$ and we will get a non-vanishing vacuum   for finite value of $\det M$. So we have a runaway potential in the $M$ direction. Adding tree level terms as a function of $M$ cannot avoid the runaway potential for $S$ in the field dependent case but may avoid the runaway potential for $M$ in the constant case, fixing $M$ at a finite value.

Another interesting case is when the matter content is $N_f=N_c$.
 In this case the second term in
eq.(\ref{W2}) vanishes and extremizing the superpotential with respect
 to $U$ gives the quantum constraint  
$\det M = \Lambda^{2N_c}$\footnote{Notice that the way this constraint
 is realized is different from the one assumed in
 reference \cite{Dvali}\ for instance, but this does not change
 any of the results of that paper.}, where we have taken the baryons v.e.v.
 to vanish. 
If   $<M_i^j> = 0$ then the quantum constraint will be satisfied only for
 a runaway dilaton field (i.e. 
vanishing $\Lambda$). This is always possible for a field dependent gauge
 coupling but it will not be satisfied 
for a finite value of the gauge coupling  or a  constant gauge coupling and, 
 therefore,   
supersymmetry will be 
broken.  Furthermore, we can 
 add to the superpotential eq.(\ref{W2})  tree level terms like $M^a +M^b$
 for the mesons   \ci{Dvali} that do 
not destroy the symmetries yielding a  finite value of $M$ and thus
 stabilizing the 
 dilaton field through the quantum constraint.

{} For $ N_f > N_c +1$ the exponent in eq.(\ref{W}) is positive. In this case, a runaway behaviour for the 
dilaton is no longer favoured since $W\rightarrow \infty$. However,  there is always a solution with 
$M \rightarrow 0$ and a  runaway of the superpotential ($W \rightarrow 0$) in the plane $S-M$ is again not  
avoided. This includes the selfdual region $\frac{3N_c}{2}<N_f<3 N_c$.

Finally consider  $3N_f > N_c  $ with all
baryons minimized by zero.  There are a number of criticisms which might be raised against
using non-asymptotically free gauge theories and against the generation of a non-perturbative superpotential 
\ci{NP}.
  However, 
the weakness in the arguments lie, in general, 
in its making an insufficient distinction between the 
effective action and the Wilson action \ci{us}.  The Wilson
action, $S_w$, describes the dynamics of the low-energy degrees
of freedom of a given system, and is used in the path
integral over these degrees of freedom in precisely the 
same way as is the classical action. The Wilson action for SQCD 
at scales for which quarks and gluons are the relevant
degrees of freedom would therefore depend on the
fields $W_{\alpha}$, ${Q^i}_\alpha$ and $\tilde{Q}_i^\alpha$.
As a result, the vanishing of $\det ( Q \tilde{Q} )$ would
indeed preclude the generation
of  a superpotential of the type $\Bigl[ e^{-8 \pi^2 S}/
\det(Q \tilde{Q})\Bigr]$ within the Wilson action.
 By contrast, it is the effective action, $\Gamma$, which is 
of interest when computing the {\it v.e.v.}s of various 
fields. And it is ${M^i}_j =  <Q^i_\alpha
\tilde {Q}_j^\alpha>$ which appears as an argument of $\Gamma$.
Since the expectation of a product of operators is not
equal to the product of the expectations of each operator,
it need not follow that $\det M = 0$ when $N_c < N_f$. 

Let us  introduce
 a mass term $Tr (\mu M) $ for the quark fields in eq.(\ref{W})  and make the mass $\mu$ dynamical, as it is  always the case  in 
string theory,  by adding a trilinear term for the field $\mu$. Eq.(\ref{W}) becomes  then
 $
W(M,\mu,S) =  Tr( \mu M)  + {h \over 3} 
\, Tr (\mu^3 )  + k \; \left( {e^{- 8 \pi^2 S} \over 
\det M } \right)^{1 /( N_c - N_f)} , 
$ 
where $k=N_c-N_f$.  
 Extremizing with respect to ${M^i}_j$, and substituting 
the result back into $W$   gives the superpotential  
 $
W(\mu,S) = {h \over 3} \, Tr \Bigl( \mu^3 \Bigr) + k' 
\Bigl( e^{- 8 \pi^2 S} \; \det \mu \Bigr)^{1 / N_c},  
$
  where $k'=N_c$. If ${\mu^i}_j$ were  
a constant mass matrix this last equation would give the 
superpotential for $S$ in SQCD. It is noteworthy that 
so long as $k' \ne 0$ the result has runaway behaviour  
to $S\rightarrow \infty$ {\it regardless} of the values of $N_c$  
and $N_f$\footnote{We thank G. Dvali for interesting discussions
 on this point.}.  We extremize, now,  $W$ with respect to the 
field ${\mu^i}_j$, to obtain the overall superpotential for 
$S$.  
The extremum is obtained for  
${\mu^i}_j=\left(-h\, e^{-8\pi^2 S/N_c}  
\right)^{N_c/(N_f-3N_c)} {\delta^i}_j$, and  the superpotential 
is then given by 
 \be 
W(S) = k'' \Bigl( h^{N_f} \; e^{24 \pi^2 S} \Bigr)^{1 /  
(N_f - 3N_c)} =k'' \Lambda^3,   
\label{NAW}\ee  
with $k''=(-1)^{3N_c/(N_f-3N_c)}\left(N_c-N_f/3\right)$.  
Notice that eq.(\ref{NAW}) takes the simple form $W \propto \Lambda^3$ 
when expressed in terms of the renormalization group invariant scale and it
 is valid for all values of $N_f$ and 
$N_c$. 
Eq.(\ref{NAW}) gives a positive exponential of $S$  if $N_f > 3 N_c$  where
 the theory is not  
asymptotically free. When this is combined with the potential 
for another, asymptotically-free  gauge group 
we obtain a superpotential of the form of eq. (\ref{W1}) with positive and
 negative exponentials 
and  a  non-trivial 
minimum can be found for $S$.
 
  The extremal condition for the dilaton $W_S = 0$ gives  
a runaway behaviour  $S\rightarrow \infty$  for
$3N_c > N_f$ but for non-asymptotically free gauge group the equation 
 $W_S=0$  
is satisfied only if  the mass field $\mu$ has a vanishing  v.e.v., i.e.
    $<\mu>=0$.  
Minimizing the superpotnetial with respect to     $\mu$, $W_\mu=0$,  gives
  two solutions: $\mu=0$ and  
$\mu=(-h e^{- 8\pi^2 S/N_c})^{3N_c-N_f}$. For asymptotically free gauge
 group $3N_c >N_f$
  both solutions 
are equivalent in the runaway limit $S\rightarrow \infty$.  In this case both 
minima are continuously connected in the $S-\mu$ plane. On the other hand, if 
$3N_c < N_f$ then the  solution  $\mu=0$ and
 $\mu=(-h e^{- 8\pi^2 S/N_c})^{3N_c-N_f}$   are driven apart 
by a large value of $S$
and one cannot continuously go from one minimum to the other one.
 The barrier between both minima
increases exponentially with increasing $S$.
  This property can play an
important  role in  the evolution of the dilaton field for cosmology.

Notice that since it is the effective
action which we use, rather than the
Wilson action, one might worry whether
our analysis is invalidated by the
appearance of nonlocal terms or
holomorphy anomalies. We argue that
this is not the case for the solution
where $\mu^i_j \ne 0$, since in this
case the matter multiplets have masses
and for scales below their mass the
theory is a pure gauge theory, which
has a gap due to confinement. Since
holomorphy anomalies arise due to
massless states,  they cannot occur
if the theory has a gap. The
same need not be true for the         
potentially runaway solution, for
which $\mu^i_j = 0$, since in this
phase there are massless matter
and gauge multiplets which can 
produce such anomalies.

An example  of dynamical global supersymmetry breaking 
with constant gauge couplings, where supersymmetry can be restored 
by the incorporation of the dilaton ($ie$ by the field 
dependence of the gauge couplings) is  the 
canonical example of dynamical global supersymmetry breaking, 
the so-called 3-2 model of Affleck {\it et al} \ci{adstwo}. In this example
the  gauge group is $SU(3)\times SU(2)$.  The fundamental matter spectrum 
is such that the $SU(2)$ factor is quantum constrained. 
  The quantum constraint 
is of the form 
$YZ=\Lambda_2^4$. 
It is shown in \cite{adsone,adstwo}\   that, if we suppose the 
condensation scale for the $SU(2)$ factor to be much 
greater than that for the $SU(3)$ factor, and if we suppose 
a certain superpotential in the microscopic 
theory, 
then the effective superpotential can be written as 
\be 
W=XY+\lambda (YZ-\Lambda_2^4)  . 
\ee 
 One can 
easily see that the equation of motion for $X$ implies 
$Y=0$ and that the equation of motion for 
the Lagrange multiplier $\lambda$ implies $YZ=\Lambda^4_2$. 
For the case of constant gauge couplings ($\Lambda_i = $constant), 
the relations cannot be simultaneously satisfied 
and supersymmetry is said to be dynamically broken. 
However, for the case of field dependent gauge couplings 
($\Lambda_i = \mu_i e^{-c_i S_i}$), the relations 
are satisfied by the runaway vacuum 
$S\rightarrow \infty$, for which $\Lambda_i=0$. 
Therefore we learn 
that in this model, supersymmetry is restored 
by a runaway dilaton if the gauge 
couplings are conceived to be field-dependent.

The opposite can also happen. We can have broken supersymmetry for field
dependent gauge coupling   but unbroken supersymmetry for field independent
 gauge coupling.
For instance, consider gaugino condensate for two gauge groups with   gauge
 kinetic function $f=f(S,T)$, 
as in string models when one-loop corrections are included. Once $T$-Duality
 is imposed on the theory the 
superpotential becomes a function of $S$ 
and $T$, $W= A_1(T)e^{-a_1 S}+A_2(T) e^{-a_2 S}$ is given in eq.(\ref{W1})
 where the $A_i$ 
coefficients
are now $T$ dependent.  It is well known that this superpotential in local supersymmetry has
a non-supersymmetric vacuum \ci{ccm}. Supersymmetry is broken through the auxiliary field of the moduli 
$T$, i.e.  
$F_T\neq 0$,  and   the dilaton gets  a finite v.e.v.   However, for a field independent   gauge  coupling, in this 
example,   supersymmetry is not broken. In global supersymmetry, there may be cases where this also happens because the field equations of the field $S$ may turn out to be inconsistent with the other field equations, but so far we have not found explicit examples showing this property.  

Finally we can also write down asymptotically free models which can
 produce positive exponentials from product group models. As an example,
 let us consider the $SU(2)\times SU(2)$ model of Intriligator, Leigh and
 Seiberg \cite{ils}\  with invariants $X$ and $Y$ for which the
 nonperturbative superpotential is:
\be
W_{np}=\frac{\Lambda_1^5\,  Y}{XY-\Lambda_2^4}
\ee
If we add to this superpotential a tree level one of the form
$W_p=\lambda_1\, X+\lambda_2 (Y-A)$ where  $\lambda_{1,2}$ are Lagrange
 multiplier fields and $A$ is a constant. We can see that integration of
 $\lambda_{1,2}$ implies $X=0, Y=A$ and so the superpotential becomes
$W_{np}=A\, \Lambda_1^4/\Lambda_2^2$. In terms of the gauge couplings
$k_1 S$ and $k_2 S$, where $k_{1,2}$ are the Kac-Moody levels of each of the two $SU(2)$ factors, 
 this is proportional to $\exp(8\pi^2(k_2-k_1)S)$ and
 therefore, for $k_2>k_1$ we have a positive exponential.
 Since $S$ is always positive this superpotential will always break supersymmetry, if we combine this model with a standard asymptotically free model we will have a sum of positive and negative exponentials and the situation will be just like the non-asymptotically free models for which $S$ can be fixed. In this case however, the limit $S\rightarrow  \infty$ is never a minimum of the scalar potential, even though we can have $W\rightarrow  0$ in this limit, and so the Dine-Seiberg general argument still holds but in a very special way, because it would imply that the runaway minimum is not continuosly connected to the finite dilaton minimum. This may also lead to very interesting cosmological features.
Notice however that the tree-level terms were just chosen to do the job, just as an illustration that this is possible.

To summarize, we have illustrated with a few examples the difference of 
considering constant gauge couplings against  field dependent gauge couplings. 
 We have  argued that the inclusion of field-dependent gauge 
couplings can qualitatively change whether or not a given model 
spontaneously breaks supersymmetry in the sense that a model with broken supersymmetry may turn out to be supersymmetric if $S$ is included. Furthermore, 
  we have found that the opposite  is also possible, i.e. 
supersymmetry can be unbroken for fixed gauge coupling, 
but  breaks down  once the gauge coupling is considered as a field.
Finally the stabilization of the dilaton field may be achieved in 
models of product groups and non asymptotically free, in a way that
may not lead to the standard runaway solution. Product group models have
 shown to provide a very rich structure and their study in more general
 cases than those considered here can lead to further surprises \cite{ustwo}.

\end{document}